\documentclass[12pt]{article}
\usepackage{html}
\usepackage{epsf}
\begin{document}
\title{MATTERS OF GRAVITY, The newsletter of the APS Topical Group on 
Gravitation}
\begin{center}
{ \Large {\bf MATTERS OF GRAVITY}}\\ 
\bigskip
\hrule
\medskip
{The newsletter of the Topical Group on Gravitation of the American Physical 
Society}\\
\medskip
{\bf Number 31 \hfill Winter 2008}
\end{center}
\begin{flushleft}
\tableofcontents
\vfill\eject
\section*{\noindent  Editor\hfill}
David Garfinkle\\
\smallskip
Department of Physics
Oakland University
Rochester, MI 48309\\
Phone: (248) 370-3411\\
Internet: 
\htmladdnormallink{\protect {\tt{garfinkl-at-oakland.edu}}}
{mailto:garfinkl@oakland.edu}\\
WWW: \htmladdnormallink
{\protect {\tt{http://www.oakland.edu/physics/physics\textunderscore people/faculty/Garfinkle.htm}}}
{http://www.oakland.edu/physics/physics_people/faculty/Garfinkle.htm}\\

\section*{\noindent  Associate Editor\hfill}
Greg Comer\\
\smallskip
Department of Physics and Center for Fluids at All Scales,\\
St. Louis University,
St. Louis, MO 63103\\
Phone: (314) 977-8432\\
Internet:
\htmladdnormallink{\protect {\tt{comergl-at-slu.edu}}}
{mailto:comergl@slu.edu}\\
WWW: \htmladdnormallink{\protect {\tt{http://www.slu.edu/colleges/AS/physics/profs/comer.html}}}
{http://www.slu.edu//colleges/AS/physics/profs/comer.html}\\
\bigskip
\hfill ISSN: 1527-3431

\begin{rawhtml}
<P>
<BR><HR><P>
\end{rawhtml}
\end{flushleft}
\pagebreak
\section*{Editorial}

The next newsletter is due September 1st.  This and all subsequent
issues will be available on the web at
\htmladdnormallink 
{\protect {\tt {http://www.oakland.edu/physics/Gravity.htm}}}
{http://www.oakland.edu/physics/Gravity.htm} 
All issues before number {\bf 28} are available at
\htmladdnormallink {\protect {\tt {http://www.phys.lsu.edu/mog}}}
{http://www.phys.lsu.edu/mog}

Any ideas for topics
that should be covered by the newsletter, should be emailed to me, or 
Greg Comer, or
the relevant correspondent.  Any comments/questions/complaints
about the newsletter should be emailed to me.

A hardcopy of the newsletter is distributed free of charge to the
members of the APS Topical Group on Gravitation upon request (the
default distribution form is via the web) to the secretary of the
Topical Group.  It is considered a lack of etiquette to ask me to mail
you hard copies of the newsletter unless you have exhausted all your
resources to get your copy otherwise.

\hfill David Garfinkle 

\bigbreak

\vspace{-0.8cm}
\parskip=0pt
\section*{Correspondents of Matters of Gravity}
\begin{itemize}
\setlength{\itemsep}{-5pt}
\setlength{\parsep}{0pt}
\item John Friedman and Kip Thorne: Relativistic Astrophysics,
\item Bei-Lok Hu: Quantum Cosmology and Related Topics
\item Gary Horowitz: Interface with Mathematical High Energy Physics and
String Theory
\item Beverly Berger: News from NSF
\item Richard Matzner: Numerical Relativity
\item Abhay Ashtekar and Ted Newman: Mathematical Relativity
\item Bernie Schutz: News From Europe
\item Lee Smolin: Quantum Gravity
\item Cliff Will: Confrontation of Theory with Experiment
\item Peter Bender: Space Experiments
\item Jens Gundlach: Laboratory Experiments
\item Warren Johnson: Resonant Mass Gravitational Wave Detectors
\item David Shoemaker: LIGO Project
\item Stan Whitcomb: Gravitational Wave detection
\item Peter Saulson and Jorge Pullin: former editors, correspondents at large.
\end{itemize}
\section*{Topical Group in Gravitation (GGR) Authorities}
Chair: Dieter Brill; Chair-Elect: 
David Garfinkle; Vice-Chair: Stan Whitcomb. 
Secretary-Treasurer: Vern Sandberg; Past Chair:  \'{E}anna Flanagan;
Delegates:
Vicky Kalogera, Steve Penn,
Alessandra Buonanno, Bob Wagoner,
Lee Lindblom, Eric Poisson.
\parskip=10pt

\vfill
\eject

\section*{\centerline
{GGR program at the APS meeting in St. Louis}}
\addtocontents{toc}{\protect\medskip}
\addtocontents{toc}{\bf GGR News:}
\addcontentsline{toc}{subsubsection}{
\it GGR program at the APS meeting in St. Louis}
\parskip=3pt
\begin{center}
David Garfinkle, Oakland University
\htmladdnormallink{garfinkl-at-oakland.edu}
{mailto:garfinkl@oakland.edu}
\end{center}
We have an exciting GGR related program at the upcoming APS April meeting
in St. Louis.\\
At the APS April meeting there will be several invited sessions of talks sponsored by the Topical Group in Gravitation (GGR).  In addition, there will be plenary talks on gravitational topics, and several of the invited sessions sponsored by other APS units are likely to be of interest to GGR members.  The invited sessions sponsored by GGR are as follows:\\ 

      Plenary Session\\

Monday, April 14, 8:30\\
Mark Kramer: The Double Pulsar: a Unique Gravity Laboratory\\
 
      GGR Invited Sessions\\ 
 
Saturday, April 12, 10:45\\ 
The Quantum Nature of Gravitational Singularities\\ 
Session Chair: Lior Burko\\ 
Beverly Berger: The nature of classical singularities\\ 
Abhay Ashtekar: Singularity resolution in loop quantum gravity\\ 
Gary Horowitz: Singularity resolution in string theory\\ 
 
Sunday, April 13, 8:30\\ 
Laboratory and Space tests of Gravitation\\ 
Session Chair: Eric Adelberger\\ 
Slava Turyshev: Pioneer Anomaly: Status of New Investigations\\ 
Jens Gundlach: Precision Test of the Equivalence Principle\\ 
Tom Murphy: Apollo: lunar laser ranging tests of gravity\\ 
 
Sunday, April 13, 13:30\\ 
Ground-based gravitational wave searches\\ 
Session Chair: Stan Whitcomb\\ 
Erik Katsavounidis: Searches for bursts of gravitational waves in LIGO, Virgo and GEO data\\ 
Duncan Brown: Gravitational waves from the inspiral of binary neutron stars and black holes\\ 
Sam Waldman: Extending our reach: plans for improved detectors in the next decade\\ 
 
Sunday, April 13, 15:30\\ 
The Broad Spectrum of Gravitation (joint with COM)\\ 
Session Chair: Pablo Laguna\\ 
Jorge Pullin: Recent Developments in Loop Quantum Gravity\\ 
Gabriela Gonzalez: The Bright Future of Gravitational Wave Astronomy\\ 
Manuel Tiglio: Numerical Relativity\\ 

\vfill\eject
 
Monday, April 14, 15:30\\ 
Astrophysics of black hole mergers (joint with DAP)\\ 
Session Chair: John Friedman\\ 
Scott Hughes: Gravitational Waves from Black Hole Mergers\\ 
Frans Pretorius: Numerical Relativity and Black hole mergers\\ 
Cole Miller: The Astrophysical Context of Black Hole Mergers\\ 
 
Tuesday, April 15, 10:45\\ 
Computational Challenges in Astrophysics, Cosmology \& Gravitation (joint with DAP)\\ 
Session Chair: Pablo Laguna\\ 
Luis Lehner: Neutron Stars in Binaries: Status and a Bright Future\\
Deirdre Shoemaker: Computing Gravity's Strongest Grip\\ 
Bronson Messer: Computational Astrophysics at the Petascale:  Toward 
Predictive Computational Science\\ 
 
Tuesday, April 15, 13:30\\ 
Numerical Relativity meets the Post Newtonian Approximation\\ 
Session Chair: Cliff Will\\ 
Emanuele Berti: Post-Newtonian diagnostics for initial data\\ 
Larry Kidder: Matching of PN waveforms with long-time numerical evolutions\\ 
Yi Pan: Developing templates from PN/Numerical waveforms\\ 

Sessions sponsored by other APS units that may be of interest to GGR
members include the following:\\

Plenary Sessions\\

Saturday, April 12, 8:30\\
Roger Blandford: Recent Developments in Plasma Astrophysics\\

Monday, April 14, 8:30\\
Michael Peskin: Dark Matter in the Cosmos and in the Laboratory\\

Invited Sessions\\

Monday, April 14, 13:30\\
Numerical Relativistic Astrophysics\\
Session Chair: Scott Hughes\\
Christian Ott: Core-collapse Supernova Mechanisms and their Signature in
Gravitational Waves\\
Alessandra Buonanno: Interplaying Analytical and Numerical Relativity in 
Modeling Binary Black Hole Coalescences\\
Fred Rasio: Hydrodynamic Calculations of Compact Binary Mergers\\

\vfill\eject

Monday, April 14, 15:30\\
Dark Energy\\
Session Chair: Michael Turner\\
Peter Garnavich: Supernovae as Probes of Dark Energy\\
Bhuvnesh Jain: Cosmological Tests of Modified Gravity vs. Dark Energy\\
Robert Nichol: Large Scale Structure and Dark Energy\\

Tuesday, April 15, 13:30\\
Short Gamma-Ray Bursts\\
Session Chair: Neil Gehrels\\
Scott Barthelmy: Observation of Prompt Emissions\\
Derek Fox: Observations of Afterglows and Hosts\\
Enrico Ramirez-Ruiz: Triggering Short Gamma-Ray Bursts\\

\vfill\eject
\section*{\centerline
{we hear that \dots}}
\addtocontents{toc}{\protect\medskip}
\addcontentsline{toc}{subsubsection}{
\it we hear that \dots , by David Garfinkle}
\parskip=3pt
\begin{center}
David Garfinkle, Oakland University
\htmladdnormallink{garfinkl-at-oakland.edu}
{mailto:garfinkl@oakland.edu}
\end{center}

Vassiliki (Vicky) Kalogera has won the APS Maria Goeppert Mayer Award. 

Francis Everitt, Eanna Flanagan, Gabriela Gonzalez, Albert Lazzarini, Lee Smolin, Massimo Cerdonio, George Gillies, Thomas Prince, and Edward Seidel have been
elected APS Fellows.

Steve Detweiler has been elected Vice-Chair of GGR

Gabriela Gonzalez has been elected Secretary-Treasurer of GGR

Larry Ford and Frans Pretorius have been elected Members at large 
of the executive committee of GGR.

Hearty Congratulations!

\section*{\centerline
{100 years ago}}
\addtocontents{toc}{\protect\medskip}
\addcontentsline{toc}{subsubsection}{
\it 100 years ago, by David Garfinkle}
\parskip=3pt
\begin{center}
David Garfinkle, Oakland University
\htmladdnormallink{garfinkl-at-oakland.edu}
{mailto:garfinkl@oakland.edu}
\end{center}
Minkowski formulated the concept of spacetime in a 1908 lecture in 
Cologne, saying in part, ``Henceforth
space by itself, and time by itself, are doomed to fade away into mere 
shadows and only a kind of union of the two will preserve an independent
reality'' (available in ``The Principle of Relativity'' Dover Publications)

\vfill\eject
\section*{\centerline{The Torsion Pendulum's Contribution to Dark Energy}}
\addtocontents{toc}{\protect\medskip}
\addtocontents{toc}{\bf Research Briefs:}
\addcontentsline{toc}{subsubsection}{\it
The Torsion Pendulum's Contribution to Dark Energy, by Bill Hamilton}
\begin{center}
Bill Hamilton, Louisiana State University
\htmladdnormallink{hamilton-at-phys.lsu.edu}
  {mailto:hamilton@phys.lsu.edu}
\end{center}

This note is motivated by the recent publication\cite{eotwash1} from the E\"ot-Wash group with new experimental constraints on the nature of gravitational potential energy.  Because of those constraints they infer limits to the size of proposed compact extra dimensions.   Central to the E\"ot-Wash experiment is a unique torsion balance. The torsion balance is frequently the instrument of choice when attempting to investigate weak forces because it can be constructed so as to have a very long period.  This is another way to say that it can be modeled as a very weak spring.  This in turn means that the instrument's deflection (angular deflection for the torsion balance) can be very large for a given applied force (torque).  This is also the curse of using the instrument since deflections will be observed from a variety of undesired sources.  The experimental challenge is to design the instrument to minimize the undesired deflections while maximizing the desired effect.

We usually think of a torsion balance as consisting of a long thin fiber supporting a dumbbell or a mass dipole.  A gravitational gradient will exert a torque on this dipole and the torque can be determined either by measuring the angular deflection of the dipole or by measuring the frequency of oscillation of the pendulum and comparing it to its frequency of small oscillation about its equilibrium position.  The modern torsion balance is mechanically more sophisticated and constructed to eliminate, as much as possible, undesired mass multipole moments\cite{eotrot,boynton} such as the dipole moment.  The E\"ot-Wash apparatus is designed to be, with the exception of small calibration masses and the mirrors used to reflect the measurement light beam, cylindrically symmetric.  The detection element is a circular disk that forms the bottom of the cylinder.  This detection disk is made sensitive to horizontal gravitational forces by drilling a pattern of holes in it (the E\"ot-Wash group thinks of these holes as negative or missing mass).  In the current experiment there are 42 holes in the detection disk, arranged in 21-fold azimuthal symmetry.  Such an arrangement in isolation is insensitive to a gravitational gradient. This statement ignores the mass balancing that must be done to eliminate the mass multipole moments arising because of imperfections in construction and assembly.

A second circular plate is drilled with 42 holes in the same pattern and aligned with and placed below and close to the detector plate.  The authors refer to this plate as the attractor plate.  If its holes are directly aligned with the holes in the detector plate it will exert no torque on the torsion balance.  However if the attractor plate is rotated about the common center axis so that the holes are displaced azimuthally its missing mass will interact with the missing mass of the detector plate and exert a torque on the balance, causing it to rotate.  Rotating the attractor plate at an angular frequency $\omega$ will, because of the 21-fold symmetry, induce a primary torque at $21\omega$ on the torsion balance. There are also components of the torque at higher harmonics of $21\omega$.

From an experimental standpoint this is usually a very desirable setup.  An input at a low frequency results in an output at a higher frequency.  In the detection system the signal at $21\omega$ will not be contaminated by pickup from the source that drives the attractor plate at frequency $\omega$.   Varying $\omega$ allows the experimenter to place the signal appropriately with respect to the free oscillation frequency of the torsion balance.  For these experiments the free oscillation period was between 500 and 650 seconds.

The description to this point describes the principal attributes of most of the recent E\"ot-Wash designs\cite{eotwash2}.  The new thing in this experiment is in the design of the lower attractor disk which is made of two circular disks, placed one on top of the other.  The second lower disk is drilled with 21 holes with twice the diameter of each of the holes on the upper disk.  The lower disk is rotated by 360/42 degrees with respect to the upper one so that the holes on the lower disk are directly below the mass between the holes on the upper disk.  The thickness of the lower disk is chosen so that the missing mass of the lower disk combined with the missing mass of the upper disk would exert practically no torque on the detector plate {\em if the inverse square law holds} (emphasis by the E\"ot-Wash group).  Thus the experiment is designed to look for violations of the inverse square law at close distances.

I must, at this point, pay homage to those who built this and the previous experiments.  A sensitive torsion balance is subject to being deflected by almost anything.  Stray electric charges, magnetic impurities, 
unmodeled gravitational potentials; all of those are capable of causing extraneous excursions of the torsion balance.  The spacing between the attractor disk and the detection disk is such that even the Casimir-Polder effect (which was only first measured in 1958 and precisely measured in 1997) has to be neutralized.

The authors parameterize\cite{eotwash2,annrev} any violation of the inverse square law by adding a Yukawa term to the Newtonian potential: 
$$V = - \frac{GM}{R} (1 + \alpha e^{-\frac{R}{\lambda}})$$ and assert that the torque would not cancel if $\lambda$ were less than the thickness of the upper disk which in this experiment is approximately 1 mm.  They convert their measured angular responses into torques (the calibration is very elegant using a rotating external gravitational octopole) and plot the measured torques and calculated torques against the separation of the attractor and detection disks and display plots of the residuals.  These residual plots must have gotten people in the lab very excited at one time.  They eventually exclude any Yukawa type interaction with $| \alpha | =1$ and $\lambda  > 56$ microns.  These results are then used to make inferences about the allowed size of compact dimensions in some of the postulated corrections to Newtonian gravity.

\section*{\centerline{10th Capra Meeting on Radiation Reaction}}
\addtocontents{toc}{\protect\medskip}
\addtocontents{toc}{\bf Conference reports:}
\addcontentsline{toc}{subsubsection}{\it  
10th Capra Meeting on Radiation Reaction, by Carlos F. Sopuerta}
\begin{center}
Carlos F. Sopuerta, Institute of Space Sciences, National Spanish Research
Council (ICE-CSIC)
\htmladdnormallink{sopuerta-at-ieec.uab.es}
  {mailto:sopuerta@ieec.uab.es}
\end{center}

The {\em Capra} meetings are devoted to the study of {\em radiation
reaction} in General Relativity, a key problem in order to understand
the general-relativistic dynamics of a two-body system.
Probably, the main application of these studies is the description
of the inspiral dynamics of stellar-mass compact objects into massive black 
holes in galactic centers, also known as extreme-mass-ratio inspirals (EMRIs), 
a primary scientific target for the future ESA/NASA Laser Interferometer Space 
Antenna (LISA) mission.  Simulating EMRIs we can obtain gravitational waveform
templates that will be crucial for the analysis of the LISA data.
The scientific outcome of these observations ranges from tests
of the {\em no-hair} theorem for black holes to the determination
of stellar populations and the study of dynamics of galactic nuclei.
In this sense, the Capra meetings and the community created around them
has contributed much of the progress in the understanding of the
radiation reaction problem and in the description of the dynamics and
gravitational-wave emission of EMRIs.

The Capra series of meetings started in August 1998 at a ranch in 
northeastern San Diego county once owned by Frank Capra, the celebrated
American film director, a Caltech alumnus who donated the ranch to Caltech,
hence the name of the meeting.  The series continued at
the following venues: UCD, Dublin (1999); Caltech, Pasadena (2000); AEI, 
Potsdam (2001); Penn State, State College (2002); YITP, Kyoto (2003); 
CGWA, Brownsville (2004); RAL, Abingdon (2005); and UWM, Milwaukee (2006).

This year, the tenth edition of the Capra Meeting and Workshop took place at 
the University of Alabama in Huntsville, from June 24th to June 29th, 2007.
It was organized by Lior Burko and was funded by the College of Science of the
University of Alabama in Huntsville. Special thanks go to Jack Fix,
and to Cindy Brasher and Dora Wynn for administrative support.
The meeting was attended by 26 researchers from research/academic 
institutions in 7 countries.  

Following the tradition of previous years,
the meeting consisted of two parts: In the first one (June 25-27) the
participants presented research talks, whereas the second one (June 27-29) 
was a 
workshop aimed at facilitating discussions, brainstorming, and collaboration. 
The program and an electronic version of all talks are available online at:

\htmladdnormallink
{http://gravity.uah.edu/capra10/program.htm}
{http://gravity.uah.edu/capra10/program.htm} \\

This year's meeting can be characterized by the amount of numerical work 
presented, especially in the time domain, which illustrates the 
progress in the field, in the sense that we are in a period in which
a substantial part of the efforts are directed towards the  
numerical implementation
of the analytical developments of past research by the Capra community.  
Another highlight of the
conference was the celebration of the 60th birthday of Steve Detweiler, 
one of the main contributors to the Capra community and, after this year's
edition, the person with most attendances at the meeting.

The meeting started with a summary by Leor Barack (Southampton) of the
progress that the Relativity group at Southampton is making in 
the development of techniques and numerical algorithms, in the time domain, for 
the computation of the self-force.   The talk served as an introduction 
to the presentations of the other Southampton group members present at the
meeting.  
The first one was Norichika Sago, who presented work on the 
numerical calculations of the gravitational self-force, in the Lorenz
gauge, for a particle orbiting a Schwarzschild black hole.  The case 
of circular orbits has been completed~\cite{Barack:2007tm} and the case of generic (eccentric)
orbits is well underway.  
Darren Golbourn, also from Southampton, gave a
talk on a new technique to regularize the self-force due to scalar-field 
perturbations from a particle orbiting a non-rotating black hole and
its numerical implementation~\cite{Barack:2007jh}.
This work is done using a 2+1 spacetime domain, which paves the way for a 
future extension to the case of a spinning Kerr black hole. The
technique proposed consists of subtracting from each azimuthal 
mode of the retarded 
field a piece that describes the singular behavior of the scalar field near the
particle.  This is done through a careful analytical analysis 
of the scalar field
near the particle, which is referred to by the authors as 
a {\em puncture scheme},
and indeed it resembles the puncture approach used in simulations
in numerical relativity.  
After the break, Ian Vega (Gainesville) presented work along the same lines as 
the previous talk but using a different approach.  In this talk the  
regularization procedure is based on the use of {\em smeared-out} sources in the 
scalar field equations, in the sense that the practical
implementation consists of applying a window function to an analytical
approximation to the singular part
of the scalar field (the approximation is constructed using THZ coordinates).
This also leads to an equation for the regular part involving an extended 
non-singular source. 
The numerical calculations are done using an evolution code in a 1+1 
spacetime domain.
Lior Burko (Hunstville) discussed progress in the time-domain extraction of 
extreme-mass-ratio binary waveforms by means of a Teukolsky numerical code
in a 2+1 domain 
and two ways of regularizing the particle sources: (i) considering a smeared-out
source based on the use of a Gaussian as an approximation to the Dirac delta
distribution~\cite{Burko:2006ua}, and (ii) using a {\em discretized} version of the Dirac delta
distribution that has a finite support in the computational domain~\cite{Sundararajan:2007jg}.
These calculations improve substantially the accuracy achieved with previous
computational schemes.   

The second day was opened by Paul Anderson (Wake Forest) with a talk on the
calculation of the self-force using the Hadamard-WKB expansion.  This is 
a completely analytic technique and the aim is to compute the non-local
contribution to the self-force (tail term) by means of a quasi-local 
expansion.  The case of a particle subject to a 
scalar field was shown as an illustration, and included a discussion of the 
range of validity of this type of expansion.
Ted Newman (Pittsburgh) presented a new approach, based on recent work~\cite{Kozameh:2006hk,Kozameh:2007sk},  
to the study of radiation reaction by looking at solutions of the
Einstein-Maxwell equations in the future null asymptotic region.
Introducing a complex worldline identified as the center of mass and charge
of the system (which were assumed to coincide), it is found that its equation 
of motion includes the known radiation-reaction terms.
Adam Pound (Guelph) discussed in the first part of his talk how to define 
adiabatic and radiative approximations to the motion of EMRIs, and what  
their ambiguities and limitations are.  In the second part, using the method
of osculating orbits in combination with a two-timescale expansion, he 
characterized the type of errors that can be produced in
the time-dependence of the orbit, in particular in the phase.  
This was illustrated in test cases of
motion under  post-Newtonian electromagnetic and gravitational self-forces~\cite{Pound:2007th,Pound:2007ti}.
Larry Price (Gainesville) reviewed recent progress on 
perturbation theory of a Kerr black hole in relation 
to the construction of Hertz potentials, discussing the existence of
radiation gauges and the question of non-radiated multipoles~\cite{Price:2006ke}.  
Bernard Whiting (Gainesville) continued with this topic and described
several tools that can help advancing in this line: GHP formalism tools, 
separability, symmetry operators, etc.
Roland Haas (Guelph) gave the last talk of the day.  He   
presented recent results on the numerical computation, using time-domain 
techniques, of the scalar self-force on eccentric orbits around a Schwarzschild 
black hole~\cite{Haas:2007rh}.  The numerical algorithm used is a fourth-order accurate 
generalization of the
{\em characteristic} code proposed initially by Lousto and Price.
He also presented a generalization of the regularization
parameters of the {\em mode-sum} regularization scheme to the case 
of non-geodesic motion.

The third day started with a talk by Carlos Sopuerta (Guelph) on 
numerical methods in the time-domain for the computation of the
gravitational perturbations generated by a particle orbiting a non-rotating
black hole.  He discussed 1+1 algorithms based on the finite element
and the pseudospectral collocation methods, showing results for 
circular orbits and discussing the prospects for the use of this method
in the case of a spinning black hole and calculations of the self-force.
Dong-Hoon Kim (Golm) gave a talk on the calculation of the self-force, via
the mode-sum scheme, on a particle moving around a slowly rotating black hole.
After the break, Bernard Whiting introduced Steve Detweiler (Gainesville)
and honored him on the occasion of his 60th birthday, emphasizing the importance
of Steve's work not only in the context of the Capra science but in the
broader context of General Relativity and its astrophysical applications.  
In his talk, Steve gave a broad picture of the gravitational self-force effects 
on orbits around a Schwarzschild black hole.  He reminded us about the main
issues that need to be tackled: Kerr metric perturbations from the Teukolsky
formalism,
search for gauge-invariant quantities, second-order perturbations generated
by a particle, etc.  He also summarized some of the progress already done
and presented during the meeting.
Jonathan Thornburg (Southampton) gave an introduction to adaptive mesh 
refinement in double-null coordinates, based on the Berger and Oliger 
scheme.  He discussed the potential applications of this numerical technique
to self-force calculations and how it can help reducing their computational cost.
The part of the meeting devoted to talks ended with a presentation by
Sam Gralla (Chicago) on the locality of the tail integral involved in
the calculation of the self-force.  His conclusion was that in the general
case this integral must be highly non-local.

The rest of the time was allocated for the workshop.   The participants
were provided with a nice atmosphere to discuss progress and collaborate.
In between discussions, and thanks to the generous hospitality of the 
U.S. Space and Rocket Center and Mr.
Cliff Broderick, the organizers and participants enjoyed a very interesting 
and instructive visit.

\section*{\centerline{Quantum Gravity in the Southern Cone IV}}
\addcontentsline{toc}{subsubsection}{\it
Quantum Gravity in the Southern Cone IV, by Rafael A. Porto}
\begin{center}
Rafael A. Porto, Department of Physics, University of California, Santa Barbara, CA 93106 
\htmladdnormallink{rporto-at-physics.ucsb.edu}
  {mailto:rporto@physics.ucsb.edu}
\end{center}

The fourth installment of the, by now traditional, `Quantum Gravity in the 
Southern Cone' meeting took place in Punta del Este, Uruguay, as it did 
eleven years ago when this series first launched. The purpose of this workshop 
was twofold. Firstly as was expected, it allowed scientists from all over 
the world to gather together, present new ideas and discuss the many 
different approaches to a quantum theory of gravity. Secondly, it provided 
an opportunity 
to celebrate the physics of Rodolfo Gambini on the 
occasion of his 60th birthday 
(although he is 61 by now).

The workshop had over 100 participants, including plenary talks by Abhay 
Ashtekar, David Berenstein, Martin Bojowald, Laurent Freidel, Juan Maldacena, 
David Mattingly, Herman Nicolai, Rafael Porto and Marcus Spradlin. There were 
afternoon sessions with talks by Jorge Alfaro, Andres Anabalon, William 
Cuervo, Alcides Garat, Horacio Girotti, Eduardo Guendelman, Olivera Miskovic, 
Maria Parisi, Albert Petrov, Michael Reisenberger, Carlos Reyes, Davi 
Rodrigues, Jose Vergara, Victor Taveras, Francesco Toppan and Luis Urrutia. 
Also Jorge Pullin reviewed Rodolfo's early work and its impact on a talk 
entitled `The Physics of Rodolfo'. In addition several participants 
contributed posters. The program and links to the talks can be found 
at 
\htmladdnormallink
{http://qgsciv.fisica.edu.uy}
{http://qgsciv.fisica.edu.uy} \\

The plenary talks covered different areas although we had three talks on the 
ADS/CFT (Anti-De Sitter/ Conformal Field Theory) correspondence and another 
three with a Loop Quantum Gravity (LQG) taste. Juan Maldacena gave an 
introduction to the ADS/CFT correspondence, the new developments and attempts 
to understand strongly coupled theories, such as QCD, as well as gluon 
scattering amplitudes. Marcus Spradlin elaborated on multiloop gluon 
amplitudes. Maldacena made two important remarks about ADS/CFT, namely 
Lorentz Invariance (LI) is preserved by quantization (in the sense that the 
dual QFT preserves conformal symmetry) and also on the issue of background 
independence, where it was pointed out that there is a sum over all possible 
geometries with ADS boundary conditions. One may argue on the necessity of a DS/CFT correspondence, or trying to implement flat boundary conditions. This however seems far from understood. During the question session the issue of whether ADS/CFT applies to QCD was brought up. Maldacena emphasized that the duality provides a sort of Ising Model as in Condensed Matter physics. One may wonder then whether we can do any better, and whether $N=4$ Super Yang-Mills theory and QCD are in the same `universality class'.
Aiming in the opposite direction to the ADS/CFT conjecture, David Berenstein talked about emergent geometry and the wave function of the universe. His talk centered on how to perform a strong coupling expansion by reducing the problem to spherically symmetric configurations in a matrix model of commuting hermitian matrices.\\ 
 
In the LQG corner, Abhay Ashtekar spoke about how quantum geometry extends 
its life past the classical singularities of GR. Ashtekar concentrated on the Big Bang, discussed Loop Quantum Cosmology (LQC), and in particular the FRW model. He also introduced the idea of  ``New Quantum Mechanics" , and that the 
holonomies are the right objects to be quantized rather than the connection. Given the so called `Bohr compactification', LQC allows for inequivalent representations (bypassing the Von Neumann theorem), and therefore avoidance of singularities unlike in the Wheeler-DeWitt approach. In LQC (relational) evolution is deterministic across the classical big bang. The main open issue in LQC is whether it can be systematically derived from full LQG due to the `gauge fixing' imposed along the way. Martin Bojowald covered LQC as well. Bojowald also talked about the effective description, where expectation values are obtained in a systematic fashion by solving a set of effective equations in the semiclassical regime, though accounting for `quantum corrections', and also its application to cosmology. Also on the LQG side Laurent Freidel  discussed spin foam models for 4d gravity. Freidel talked about the topological BF theory and its quantization, how to implement the simplicity constraints of gravity, and how the resulting spin foam model resembles the kinematical states (spin networks) in LQG. Yet is there to understand the link with the dynamics.\\

Among the plenary speakers we also had Hermann Nicolai talking about `Re-inventing M theory'. Nicolai talked about $E_{10}$, and how it incorporates many relations between maximal supergravity theories. Victor Rivelles talked about the S-matrix of the Faddeev-Reshetikhin model and its connection with the ADS/CFT correspondence. Jorge Zanelli, who spoke about Chern-Simons gravity and the Universe as a topological defect, and Maximo Banados on GR with $g_{\mu\nu}=0$ as the `ground state', where diffeomorphisms act as isometries. Banados elaborated on a would-be theory of gravity which incorporates the `no metric' condition and argued this theory could provide candidates for dark energy and dark matter. David Mattingly talked about violations of LI (LIV) and QG phenomenology. Experimental constraints on LIV are very tight. Also, even if one argues that violations come from higher dimensional operators, $d \le 4$ operators can be generated by radiative corrections. That introduces fine tunning, which one may argue could be avoided if CPT is unbroken, and SUSY is implemented, though broken at the TeV scale. 
From a QG point of view, LIV are highly suppressed, and that represents a strong constraint on the theories.

Summarizing, the meeting was very fruitful and interactive, with many discussions and exchange of ideas taking place. The next meeting will take place in Brasil.


\begin{thebibliography}{9}

\bibitem{eotwash1} 
D.J. Kapner, T.S. Cook, E.G. Adelberger, J.H. Gundlach, B.R. Heckel, C.D. Hoyle, and H.E. Swanson, Phys. Rev. Lett. {\bf 98}, 021101 (2007).

\bibitem{eotrot}
G.L. Smith, C.D. Hoyle, J.H. Gundlach, E.G. Adelberger, B.R. Heckel, and H.E. Swanson, Phys. Rev. D {\bf 61}, 022001 (2000)

\bibitem{boynton}
P.E. Boynton, R.M. Bonicalzi, A.M. Kalet, A.M. Kleczewski, J.K. Lingwood, K.J. McKenney, M.W. Moore, J.H. Steffen, E.C. Berg, W.D. Cross, R.D. Newman, and R.E. Gephart, New Astronomy Reviews {\bf 51},  334 (2007).

\bibitem{eotwash2}
C.D. Hoyle, D.J. Kapner, B.R. Heckel, E.G. Adelberger, J.H. Gundlach, U. Schmidt, and H.E. Swanson, Phys. Rev. D {\bf 70}, 042004 (2004).

\bibitem{annrev}
E.G. Adelberger, B.R. Heckel, and A.E. Nelson, Annu. Rev. Nucl. Part. Sci. 2003. {\bf 53}:77-121.


 
\end{thebibliography}

\begin{thebibliography}{9}
\bibitem{Barack:2007tm}
  L.~Barack and N.~Sago,
  Phys.\ Rev.\  D {\bf 75}, 064021 (2007)
\htmladdnormallink{[arXiv:gr-qc/0701069].}{http://arxiv.org/abs/gr-qc/0701069}
  
\bibitem{Barack:2007jh}
  L.~Barack and D.~A.~Golbourn,
  Phys.\ Rev.\  D {\bf 76}, 044020 (2007)
\htmladdnormallink{[arXiv:0705.3620 [gr-qc]].}{http://arxiv.org/abs/0705.3620}
  
\bibitem{Burko:2006ua}
  L.~M.~Burko and G.~Khanna,
  Europhys.\ Lett.\  {\bf 78}, 60005 (2007)
\htmladdnormallink{[arXiv:gr-qc/0609002].}{http://arxiv.org/abs/gr-qc/0609002}

\bibitem{Sundararajan:2007jg}
  P.~A.~Sundararajan, G.~Khanna and S.~A.~Hughes,
\htmladdnormallink{arXiv:gr-qc/0703028.}{http://arxiv.org/abs/gr-qc/0703028}
  
\bibitem{Kozameh:2006hk}
  C.~Kozameh, E.~T.~Newman and G.~Silva-Ortigoza,
  Class.\ Quant.\ Grav.\  {\bf 23}, 6599 (2006)
\htmladdnormallink{[arXiv:gr-qc/0607074].}{http://arxiv.org/abs/gr-qc/0607074}

\bibitem{Kozameh:2007sk}
  C.~Kozameh, E.~T.~Newman and G.~Silva-Ortigoza,
\htmladdnormallink{arXiv:0706.2318 [gr-qc].}{http://arxiv.org/abs/0706.2318}
  
\bibitem{Pound:2007th}
  A.~Pound and E.~Poisson,
\htmladdnormallink{arXiv:0708.3033 [gr-qc].}{http://arxiv.org/abs/0708.3033}

\bibitem{Pound:2007ti}
  A.~Pound and E.~Poisson,
\htmladdnormallink{arXiv:0708.3037 [gr-qc].}{http://arxiv.org/abs/0708.3037}

\bibitem{Price:2006ke}
  L.~R.~Price, K.~Shankar and B.~F.~Whiting,
  Class.\ Quant.\ Grav.\  {\bf 24}, 2367 (2007)
\htmladdnormallink{[arXiv:gr-qc/0611070].}{http://arxiv.org/abs/gr-qc/0611070}
  
\bibitem{Haas:2007rh}
  R.~Haas,
  Phys.\ Rev.\  D {\bf 75}, 124011 (2007)
\htmladdnormallink{[arXiv:0704.0797v2 [gr-qc]].}{http://arxiv.org/abs/0704.0797}
  
\end{thebibliography}
\end{document}